\documentclass[aps,twocolumn,prl,superscriptaddress]{revtex4}
\usepackage[dvips]{graphics}
\usepackage{graphicx}
\usepackage{amsfonts}
\usepackage{amssymb}
\usepackage{amsmath}
\usepackage{subfigure}

\begin{document}

\title{Nonlinear susceptibilities and the measurement of a cooperative length}
\author{E. Lippiello}
\affiliation{Dipartimento di Scienze Fisiche, Universit\'a di Napoli
``Federico II'', 80125 Napoli, Italy}
\author{F.Corberi}
\affiliation{Dipartimento di Matematica ed Informatica,
via Ponte don Melillo, Universit\`a di Salerno, 84084 Fisciano (SA), Italy}
\author{A.Sarracino}
\affiliation{Dipartimento di Fisica ``E.R.Caianiello'',
via S.Allende, Universit\`a di Salerno, 84081 Baronissi (SA), Italy}
\author{M. Zannetti}
\affiliation{Dipartimento di Matematica ed Informatica,
via Ponte don Melillo, Universit\`a di Salerno, 84084 Fisciano (SA), Italy}

 \date{\today}

\begin{abstract}
We derive the exact beyond-linear fluctuation dissipation relation, connecting the response of
a generic observable to the appropriate correlation functions, 
for Markov systems. The relation, which takes a similar form for systems governed by a 
master equation or by a Langevin equation, can be derived to every order, in large generality 
with respect to the considered model, in equilibrium
and out of equilibrium as well. On the basis of the fluctuation dissipation
relation we propose a particular response function, namely the second order
susceptibility of the two-particle correlation function, as an effective quantity to detect 
and quantify cooperative effects in glasses and disordered systems. We test this
idea by numerical simulations of the Edwards-Anderson model in one and two dimensions.
\end{abstract}

\pacs{05.70.Ln, 75.40.Gb, 05.40.-a}

\maketitle

\draft
\def\be{\begin{equation}}
\def\ee{\end{equation}}
\def\bfi{\begin{figure}}
\def\efi{\end{figure}}
\def\bea{\begin{eqnarray}}
\def\eea{\end{eqnarray}}
\newcommand{\ket}[1]{\vert#1\rangle}
\newcommand{\bra}[1]{\langle#1\vert}
\newcommand{\braket}[2]{\langle #1 \vert #2 \rangle}
\newcommand{\ketbra}[2]{\vert #1 \rangle  \langle #2 \vert} 

A central phenomenon in the statistical mechanics of interacting systems is 
the onset of long range order when approaching phase transitions,
specifically second order ones such as the para-ferromagnetic or gas-liquid transition.
The coherence length $\xi$ expressing the range of correlations is disclosed
by the knowledge of an appropriate (two point) correlation function $C_{ij}$, as is 
$C_{ij}=\langle \sigma _i \sigma _j \rangle-
\langle \sigma _i \rangle \langle \sigma _j \rangle$ for the prototypical Ising model.
The divergence of $\xi $ induces the scaling symmetry  
when the critical point is neared. 
In this framework, equilibrium linear response theory, relating $C_{ij}$
to its conjugate susceptibility $\chi _{ij}$ 
(and more generally two time correlations 
$C_{ij}(t_1,t_2)=\langle \sigma _i (t_1)\sigma _j (t_2)\rangle
-\langle \sigma _i (t_1)\rangle \langle \sigma _j (t_2)\rangle$ and susceptibilities 
$\chi _{ij}(t_1,t_2)$)
through the fluctuation-dissipation theorem (FDT), 
has proved to be of the uppermost importance both theoretically
and experimentally, allowing the alternative determination of correlations,
and hence of $\xi $, through linear response functions. 

These concepts are not restricted only to equilibrium states, 
but inform non-equilibrium statistical mechanics as well.
For example, in a broad class of aging systems the kinetics is characterized by 
the growth of a characteristic length $L(t)$, determining 
a dynamical scaling symmetry in close analogy
to what happens in static phase-transitions.
In view of these and related issues, increasing interest has been recently devoted to 
the generalization of linear response theory to out of equilibrium systems,
a research subject originating from
the recognition that the relation between $\chi _{ij}(t_1,t_2)$ and $C_{ij}(t_1,t_2)$
may be used to define an {\it effective} temperature~\cite{teff} and to
bridge between equilibrium and non equilibrium properties~\cite{fmpp}.
Although a theorem of such a generality as the FDT cannot be derived off equilibrium, 
in the case of Markov processes a natural generalization in the form of a 
fluctuation-dissipation relation (FDR)
between $\chi _{ij}(t_1,t_2)$, $C_{ij}(t_1,t_2)$ and a correlator $D_{ij}(t_1,t_2)$
involving the generator of the stochastic process has been
obtained~\cite{CKP,rlin}. 
This result could open the way, in principle, to measurements of 
$C_{ij}(t_1,t_2)$, and hence of $L(t)$, from non equilibrium susceptibilities,
provided the properties of $D_{ij}$ are known. 

This whole approach cannot be straightforwardly applied to the case of
glasses, spin glasses and in several instances of disordered systems,
because their unusual type of long range order is not captured by  
linear response functions or even by two point correlators: These quantities 
remain short-ranged, even when some long range order appears in the system.
This is because ordered patterns are randomized by the quenched disorder
so that, for instance, $\overline {\langle \sigma _i \sigma _j\rangle}$
(where the overbar denotes the average over the disorder)
vanishes even when $\langle \sigma _i \sigma _j\rangle \neq 0$.         
To circumvent this problem, one has to consider higher order (non linear)
response functions or, equivalently, $n$-spin ($n > 2$) correlation functions $C^{(n)}$.
Along this line, recently, a measure of cooperativity 
has been proposed~\cite{8dibouchaudbiroli}
relying on a four point correlation function as
\begin{eqnarray}
C^{(4)}_{ij}(t,t_w) & = &
\overline {\langle \sigma _i(t) \sigma _i(t_w) \sigma _j(t) \sigma _j(t_w)\rangle } \nonumber \\
& - & \overline {\langle \sigma _i(t) \sigma _i(t_w)\rangle }
\overline {\langle \sigma _j(t) \sigma _j(t_w)\rangle}.
\label{c4} 
\end{eqnarray}
The idea is that, while $C_{ij}$ is annihilated by 
the disorder average, 
the variance of $\sigma _i \sigma _j$ survives, possibly providing informations
on cooperativity.
$C^{(4)}_{ij}$ has been proved to be
effective in numerical simulations~\cite{simu,mayer} but
its direct experimental investigation remains a challenge~\cite{mayer}, 
as in general multi-point correlators.
A natural way out of this deadlock is to measure responses to
external perturbations, namely susceptibilities,
as suggested by Bouchaud and Biroli~\cite{biroli} and
done experimentally in~\cite{science}. 
In order to make sure what actually do the non-linear susceptibilities probe,
however, it is crucial to establish their relationship with multi-point correlators. 
Some specific aspects
of this issue have been considered recently~\cite{semer,biroli},
limited to the case of systems governed by a Langevin equation,
but a general formulation is presently lacking.
 
In this Paper, we present the exact derivation of the FDR beyond linear order
for spin models evolving with Markovian dynamics. The systematic approach we use
is quite general, allowing one to derive the response function of an arbitrary 
observable to every order in the external perturbation and to relate it to correlation functions 
of the unperturbed system,
in equilibrium and out of equilibrium as well, for generic spin
models (e.g. Ising, clock, Heisenberg models etc ...) in full
generality with respect to the Hamiltonian and the evolution rules.
We show that the FDR takes the same form for hard spins,
whose kinetics is ruled by a master equation, and 
for soft spins systems governed by a Langevin equation, further supporting the
generality of our result. This relation shows that, already in equilibrium,
beyond linear order the susceptibility is related not only to multi-spin correlations $C^{(n)}$ 
but also to the $D$ correlators,
much like in linear theory out of equilibrium. This feature loosens
the relation between response and multi-spin correlations, raising the
question of which response function is best suited to detect
cooperative effects.
We argue that a particular susceptibility $\chi ^{(c,2)}$,
basically the second order response of the correlation function $C$,  
is well fit to this task, and bears informations on the correlation length. 
We complement this idea by numerical simulations of disordered
spin models, showing how the existence of a growing length 
can be detected using $\chi ^{(c,2)}$.

Let us sketch the derivation of the FDR for hard spins~\cite{nota2}. 
Using the operator formalism,  
we consider for simplicity a system of Ising spins (but the result holds more generally) 
whose state is described
by the vector $\ket {\sigma}=\bigotimes \ket{\sigma_i}$ ($i=1,N$) on a lattice. 
The stochastic evolution is characterized by the propagator 
\be
\hat{P}(t|t_w)= {\cal T} \exp \left (\int_{t_w}^t ds \hat{W}(s) \right),
\label{propagator}
\ee
where $\hat{W}(t)$ is the time dependent generator of the process, which is assumed to 
obey detailed balance, and ${\cal T}$ is the time ordering operator. 
The expectation  $\langle {\cal O} (t)\rangle $
of a generic observable ${\cal O} $ on the time dependent state 
$\vert P(t)\rangle$ is given by $\langle -\vert \hat {\cal O} \vert P(t)\rangle$, 
where $\bra{-} = \sum_{\sigma}\bra{\sigma}$ is the flat vector. 
Using the propagation $\ket{P(t)}= \hat{P}(t| t_w)\ket{P(t_w)}$ of the states 
this can be written as $\langle -\vert \hat {\cal O} \hat P(t\vert t_w)\vert P(t_w)\rangle$.
Switching on an external field $h$ (perturbation) at time $t_w$,
changing $\hat{P}$ to $\hat{P}_h$, the expectation $\langle {\cal O} (t)\rangle _h 
=\langle -\vert \hat {\cal O} \hat P _h(t\vert t_w) \vert P(t_w)\rangle$ can be 
expanded as $\langle  {\cal O} (t) \rangle_h  = \langle  {\cal O} (t) \rangle_0 
+\sum_{n=1}^{\infty} (1/ n!)\sum_{j_1...j_n}
\int_{t_w}^t dt_1 ...\int_{t_w}^t dt_n \; R^{({\cal O},n)}_{j_1...j_n}(t,t_1,...,t_n)$
$h_{j_1}(t_1)...h_{j_n}(t_n)$, where
\begin{eqnarray}
& & R^{({\cal O},n)}_{j_1...j_n}(t,t_1,...,t_n) = \left . {\delta^n \langle {\cal O} (t) \rangle_h \over
\delta h_{j_1}(t_1)... \delta h_{j_n}(t_n)} \right |_{h=0} \nonumber \\
& = & \bra{-} {{\cal O}} \left . {\delta^n \hat{P}_h(t| t_w) \over
\delta h_{j_1}(t_1)... \delta h_{j_n}(t_n)} \right |_{h=0} \ket{P(t_w)} 
\label {RF3}
\end{eqnarray}
is the $n$-th order response function ($t\geq t_1,...,t_n$). 
Let us workout $R^{({\cal O},2)}$ as an illustration,
the generalization to arbitrary $n$ being straightforward~\cite{nota2}. 
From~(\ref{propagator}) one has
\begin{eqnarray}
{\delta^2 \hat{P}_h(t| t_w) \over \delta h_{j_1}(t_1) \delta h_{j_2}(t_2)} &=&  
\hat{P}_h(t| t_1) 
{\partial \hat{W}(t_1) \over \partial h_{j_1}(t_1)} \hat{P}_h(t_1| t_2) \nonumber \\
{\partial \hat{W}(t_2) \over \partial h_{j_2}(t_2)}\hat{P}_h(t_2| t_w) 
&+& \hat{P}_h(t| t_1){\partial^2 \hat{W}(t_1) \over \partial h_{j_1}^2(t_1)} \hat{P}_h(t_1| t_w)
\delta _{12}
\label{RF.02}
\end{eqnarray}
 where $t_1\geq t_2$ and
$\delta _{12} = \delta_{j_1,j_2}\delta(t_1-t_2)$. We choose a perturbation 
entering the Hamiltonian as $-\sum _i h_i(t)\hat \sigma _i ^z$, 
where $\hat \sigma ^z$ is the $z$ Pauli matrix.
Assuming single spin flip dynamics for simplicity, the generalization
to multiple spin flips being straightforward,
the derivative of the generator is 
$\partial^n \hat{W}(t_1)/\partial h_{j_1}^n(t_1) =
(-\beta)^n \hat{W}_{j_1}(t_1)(\hat{\sigma}^z_{j_1})^n.$
Then
\begin{eqnarray}
& & R^{({\cal O},2)}_{j_1j_2}(t,t_1,t_2)  =  \nonumber \\
& & \beta^2 \bra{-} \hat{\cal O} 
\hat{P}(t| t_1)\hat{W }_{j_1}\hat{\sigma}^z_{j_1}
\hat{P}(t_1| t_2)\hat{W }_{j_2}\hat{\sigma}^z_{j_2}\ket{P(t_2)} \nonumber \\
& + & \beta^2 \bra{-} \hat{\cal O} \hat{P}(t| t_2)
\hat{W }_{j_2}\ket{P(t_2)}
\delta_{12}. 
\label{NL3} 
\end{eqnarray}
In order to obtain an expression involving only observable quantities
(i.e. diagonal operators), we write
$\hat{W }_{j_1}\hat{\sigma}^z_{j_1}= {1 \over 2}[\hat{W }_{j_1},\hat{\sigma}^z_{j_1}] + 
{1 \over 2} \{ \hat{W }_{j_1},\hat{\sigma}^z_{j_1} \}$,
where $[\cdot ]$ or $\{ \cdot \}$ denote the commutator or the anticommutator.
It can be easily shown that
$\hat{B}_i(t) = \{\hat{\sigma}^z_i,\hat{W}_i(t)\}$ 
is a diagonal operator  
with the property ${\partial \over \partial t} \langle 
{\sigma}^z_i(t) \rangle = \langle {B}_i (t) \rangle $.
Since the term with the commutator acts like a time derivative, 
the second order FDR is obtained
\begin{eqnarray}
&&R^{({\cal O},2)}_{j_1j_2}(t,t_1,t_2) = \frac{\beta ^2}{4} 
\Big \{ {\partial \over \partial t_1}{\partial \over \partial t_2}
\langle  {\cal O} (t){\sigma}_{j_1}(t_1)
{\sigma}_{j_2}(t_2)\rangle  \nonumber \\
&-&{\partial \over \partial t_1}
\langle  {\cal O} (t){\sigma}_{j_1}(t_1)
{B}_{j_2}(t_2)\rangle 
-{\partial \over \partial t_2}
\langle {\cal O} (t){B}_{j_1}(t_1)
{\sigma}_{j_2}(t_2)\rangle \nonumber \\  
&+& \langle  {\cal O} (t){B}_{j_1}(t_1)
{B}_{j_2}(t_2)\rangle  \Big \}
+\frac{\beta^2}{2} \langle {\cal O} (t) \sigma_{j_1}(t_1)B_{j_1}(t_2) \rangle \nonumber \\   
&\times & \delta_{j_2,j_1} \delta(t_1 - t_2).
\label{fdr} 
\end{eqnarray}
Care must be used for $t_2\to t_1$ since the product of the commutators
generates a singular term~\cite{nota2}. 
In a stationary state, using Onsager reciprocity, the above result
simplifies to
\begin{eqnarray}
&&R^{({\cal O},2)}_{j_1j_2}(t,t_1,t_2)=\frac{\beta^2}{2} 
\left \{ {\partial \over \partial t_1}{\partial \over \partial t_2}
\langle {\cal O} (t){\sigma}_{j_1}(t_1)
{\sigma}_{j_2}(t_2)\rangle  \right . \nonumber \\
& - & \left . {\partial \over \partial t_2}
\langle {\cal O} (t){B}_{j_1}(t_1)
{\sigma}_{j_2}(t_2)\rangle \right \} 
+\frac{\beta ^2}{2} \langle {\cal O}(t) \sigma _{j_1}(t_1)B_{j_1}(t_2)\rangle \nonumber \\
&\times &\delta_{j_2,j_1} \delta(t_1 - t_2).
\label{fdreq} 
\end{eqnarray}
Let us mention that for continuous variables (soft spins) governed by a Langevin equation
$\partial \sigma _i(t)/\partial t=B_i(t)+\eta _i(t)$, by taking 
$\hat W$ as the Fokker-Planck generator, we obtain~\cite{nota5}
the same FDR~(\ref{fdr}) (and hence~(\ref{fdreq})), 
without the last term containing the 
$\delta$-functions. Since on the r.h.s. do only appear correlation functions of 
the unperturbed system, Eq.~(\ref{fdreq}) qualifies as the beyond-linear FDT,
while Eq.~(\ref{fdr}) as its non-equilibrium generalization.
This relation can be derived for the response of an arbitrary 
observable to every order in the external perturbation, for hard and soft spins
alike, without reference to a particular Hamiltonian or transition rates.
Exactly like in the linear case~\cite{rlin}, the above FDR serves as the basis for
the development of a no-field algorithm for the fast computation of the non linear response
function, as it will be shown below.

The peculiar feature of the non-linear FDR~(\ref{fdr},\ref{fdreq}) is the ubiquitous 
(even in equilibrium) presence of the correlators $D$ containing the operator $\hat B$, 
which introduces a
specific reference to the particular dynamical process through the 
generator. This hinders a direct relation between response and multi-spin
correlation functions, hampering the procedure to associate
$\xi $ to a susceptibility, as in equilibrium linear theory.   
Despite this, we argue that a quantity related to the second order response of the
composite operator
$\hat {\cal O} =\hat c_{ij}=\hat \sigma ^z_i \hat \sigma ^z_j$ 
\begin{eqnarray}
-{\cal R}^{(c,2)}_{ij}(t,t_1,t_2) & = &  
\left .\frac {\delta ^2\langle \sigma _i (t)\sigma _j(t)\rangle}
{\delta h_i(t_1)\delta h_j(t_2)}\right | _ {h=0} \nonumber \\
& - & R^{(\sigma ,1)}_{ii}(t,t_1) R^{(\sigma ,1)}_{jj}(t,t_2),
\label{calr2}
\end{eqnarray}
where $R^{(\sigma ,1)}_{ij}(t,t_1)$
is the linear response function of the spin $\sigma _i$~\cite{rlin},
or, alternatively, the susceptibility
\be
\chi ^{(c,2)}_{ij}(t,t_w)=\int _{t_w}^t dt_1 \int _{t_w}^t dt_2 {\cal R}^{(c,2)}_{ij}(t,t_1,t_2),
\label{calchi2}
\ee
is well suited to detect cooperative effects (for disordered systems
a disorder average is implicitly assumed), and may be used to determine $\xi $. 
In equilibrium systems this is readily seen, since   
a simple statistical mechanical calculation yields
\be
\chi ^{(c,2)}_{ij,eq}=
\lim _{t\to \infty}\chi ^{(c,2)}_{ij}(t,t_w)= 
\beta ^2\lim _{t\to \infty}[C _{ij}(t,t)]^2=\beta ^2 C^2 _{ij,eq},
\label{chieq}
\ee
namely the counterpart of the standard static equilibrium relation
between correlations and susceptibilities. Taking the $k=0$ component 
$\chi ^{(c,2)}_{k=0,eq}=(1/N)\sum _{i,j}\chi ^{(c,2)}_{ij,eq}\propto \xi ^{4-d-2\eta}$,
therefore, one has direct access to the coherence length. 
Concerning the full two-time dependence of $\chi ^{(c,2)}$, in a system characterized 
by dynamical scaling, by virtue of Eq.~(\ref{chieq}) 
one expects the same scaling form, with the same exponents,
of $C^2$, hence 
\be
\chi ^{(c,2)}_{k=0}(t,t_w)=\xi ^{4-d-2\eta}f\left (\frac {\xi}{L(t)},
\frac{L(t_w)}{L(t)} \right ). 
\label{chiscal}
\ee
On physical grounds, one may understand why  
cooperativity effects are revealed by $\chi ^{(c,2)}$ as follows:
writing the susceptibility 
$\chi^{(\sigma ,1)}_{ij}(t,t_w)=\int _{t_w}^t dt_1 R^{(\sigma ,1)}_{ij}(t,t_1)$ 
as $\chi^{(\sigma ,1)}_{ij}(t,t_w)=\langle x_{ij}(t,t_w)\rangle$,
where~\cite{rlin} $ x_{ij}(t,t_w)={\beta \over 2} \left [
{\sigma}_i (t){\sigma}_j (t)- {\sigma}_i (t)
{\sigma}_{j}(t_w) -{\sigma}_i (t) \int _{t_w}^t  dt_1{B}_{j}(t_1) 
\right ]$,
in view of Eq.~(\ref{NL3}), $\chi ^{(c,2)}$ 
can be cast as 
$-\chi^{(c,2)}_{ij}(t,t_w)= 
\langle x^{(\sigma,1)}_{ii}(t,t_w) x^{(\sigma,1)}_{jj}(t,t_w)\rangle-
\langle x^{(\sigma,1)}_{ii}(t,t_w)\rangle \langle x^{(\sigma,1)}_{jj}(t,t_w)\rangle$.
Namely, $\chi ^{(c,2)}$ is the correlation of the variable whose
average  yields $\chi^{(\sigma ,1)}$, much in the same way as
$C^{(4)}_{ij}(t,t_w)$ is the correlation of the variable 
$\sigma _i(t) \sigma _i(t_w)$ whose average
gives $C$. Since $\chi^{(\sigma ,1)}$ is the response function
conjugated to $C$ by the FDT, this suggests
that $\chi^{(c,2)}$ may be suitable (as will be
further shown numerically below), to study
cooperativity analogously,
and for the same mechanism of $C^{(4)}$.
Despite this, $\chi^{(c,2)}$ and $C^{(4)}$
can hardly be
related. Actually, although $C^{(4)}$ appears in the first term on the r.h.s.
of the FDR~(\ref{fdr},\ref{fdreq}) for ${\cal R}^{(c,2)}$, the terms containing
$B$ spoil the relation between ${\cal R}^{(c,2)}$ and $C^{(4)}$. 
It can be shown, in fact, that in most cases
these terms are comparable with the first. For example, the static 
relation~(\ref{chieq}) depends crucially on the contributions of the terms 
containing $B$. 
 
An important advantage of $\chi ^{(c,2)}$ with respect to multi-spin correlations
is its fitting to experimental measurements.  
In fact, switching on a field $h_i$ from $t_w$ 
onwards one has $\langle \sigma _i (t) \sigma _j(t) \rangle _h=
\langle \sigma _i (t) \sigma _j(t) \rangle _{h=0}+ 
\sum _{l,m}h_l h_m \int _{t_w}^tdt_1 \int _{t_w}^tdt_2 
\delta ^2 \langle \sigma _i (t) \sigma _j(t) \rangle /
(\delta h_l (t_1)\delta h_m(t_2))+ O(h^4)$. In disordered systems the first term
on the r.h.s. vanishes and the only non-vanishing terms in the sum are those with 
$l=i$ and $m=j$ (or $l=j$ and $m=i$). Hence, using the 
definitions~(\ref{calchi2},\ref{calr2},\ref{RF3})
$\langle \sigma _i (t) \sigma _j(t) \rangle _h-\langle \sigma _i (t)\rangle ^2_h=
-h_ih_j\chi ^{(c,2)} _{ij}(t,t_w) + O(h^4)$. Therefore, the determination of 
$\chi ^{(c,2)}$ can be reduced to the measurement of a correlation 
function in an external field (for instance a uniform one).  

\begin{figure}
    \centering
    
   \rotatebox{0}{\resizebox{.45\textwidth}{!}{\includegraphics{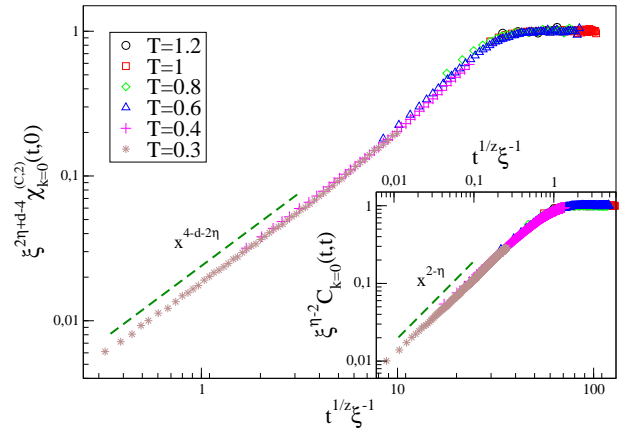}}}
    \caption{Data collapse of $\chi ^{(c,2)}$ ($C$ in the inset) for several temperatures
      in the $d=1$ EA model. The dashed lines are the expected power-laws in the non equilibrium regime.}
\label{d1}
\end{figure}

\begin{figure}
    \centering
    
   \rotatebox{0}{\resizebox{.45\textwidth}{!}{\includegraphics{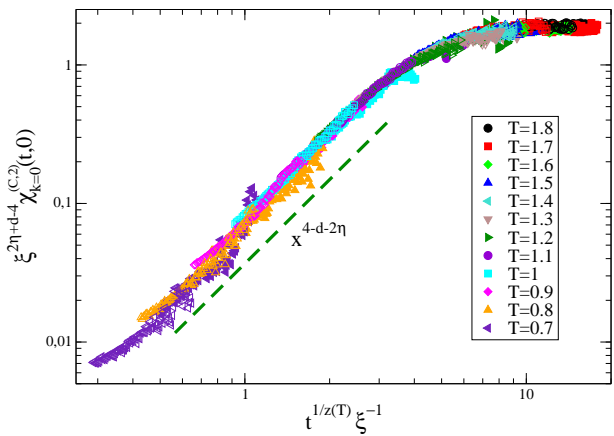}}}
    \caption{Data collapse of $\chi ^{(c,2)}$ 
for several $T$ in the $d=2$ EA model with  
bimodal (open symbols)or Gaussian (filled symbols) bond distribution,
with $z(T)=4/T$. 
The dashed line is the expected power-law in the non equilibrium regime.}
\label{d2}
\end{figure}

In order to check these ideas and to test the efficiency of the 
method to measure the cooperative length 
we have computed numerically $\chi ^{(c,2)}_{k=0}(t,0)$ in the 
Edwards-Anderson (EA) model with Hamiltonian $H=\sum _{ij} J_{ij} \sigma _i \sigma _j$
in $d\le 2$, simulated by means of standard Montecarlo techniques,
with Glauber transition rates, where $B_i=\sigma _i-\tanh(\beta \sum _{j} J_{ij}\sigma _j)$.
The system is quenched 
from a disordered state at $t=0$ to different final temperatures $T>0$.  
$\chi ^{(c,2)}_{k=0}(t,0)$ is computed using Eq.~(\ref{fdr}).
It must be stressed that, due to the noisy nature of response functions, 
the advantage provided by the FDR~(\ref{fdr}) instead of applying
an infinitesimal perturbation is numerically un-renounceable.
In fact, besides providing an incomparably better signal/noise ratio, 
the $h\to 0$ limit is built in the FDR. 
The analysis of the data proceeds as follows:
from the large $t$ value $\chi ^{(c,2)}_{k=0,eq}$ of $\chi ^{(c,2)}$, 
knowing $\eta $, $\xi $ can be extracted for each temperature.
Regarding $L(t)$,
in the non-equilibrium regime $L(t)\ll \xi $, $\chi ^{(c,2)}$ must be
independent from $\xi $. Using~(\ref{chiscal}) this implies 
$f(\xi/L(t),0) \sim (L(t)/\xi)^{4-d-2\eta}$. Hence
the non-equilibrium behavior of $L(t)$ can also be determined. 
With these results, 
one can control that data collapse is obtained by plotting 
$\xi^{-4+d+2\eta}\chi ^{(c,2)}_{k=0}(t,0)$ vs $L(t)/\xi$ for all the temperatures
considered (see Figs.~\ref{d1},\ref{d2}). 
We have studied first 
the model in $d=1$ with bimodal distribution of the coupling constants $J_{ij}=\pm 1$.  
This system can be considered as a laboratory since 
it can be mapped onto a ferromagnetic system where $\eta =1$ and $L(t)\sim t^{1/z}$,
with $z=2$, are known analytically. Moreover, besides $\chi ^{(c,2)}$, one can also check the scaling 
of the usual correlation $C_{k=0}(t,t)$ after the mapping and
obtain another determination of $L(t)$ and $\xi $. 
In doing so, we find that the two methods to extract $L(t)$ and $\xi$ 
agree within the numerical uncertainty between them, and with the analytical behaviors.
The data collapse of $\chi ^{(c,2)}$ and of $C$ is shown in Fig.~\ref{d1}.
Here one clearly observes the non-equilibrium kinetics in the early regime,
characterized by a power-law behavior of $\chi ^{(c,2)}$ with exponent $4-d-2\eta$, as expected,
and the late equilibration with the convergence of 
$\chi ^{(c,2)}_{k=0}(0,t)$ to $\chi ^{(c,2)}_{k=0,eq}$.
$C$ behaves similarly.
After this explicit verification, we turn to the $d=2$ case,
where the reference to $C$ is not available.
In this case, with  both bimodal and Gaussian distributions of $J_{ij}$,
using $\eta =0$~\cite{sgd2bis}, we find a behavior of $\xi $
consistent with previous results~\cite{sgd2bis, sgd2}.
The non-equilibrium behavior is compatible with a power law
$L(t)\sim t^{1/z(T)}$ with a temperature dependent exponent in agreement
with $z(T)\simeq 4/T$, as reported in~\cite{sgd2z}. 
The data collapse of $\chi ^{(c,2)}$ is shown in Fig.~\ref{d2}. Notice also the 
additional collapse of the curves with bimodal and Gaussian bond distribution,
further suggesting that the two models may share the same universality class
at finite temperatures~\cite{sgd2bis}. 

In this Paper we have derived the exact beyond-linear FDR. The result, which can be
straightforwardly extended to every order, provides a rather general relation
between response and correlation functions: It is satisfied by systems 
described by a master equation or by a Langevin equation, without 
reference to  specific aspects of the considered model. On the basis of the FDR we argued, 
providing numerical evidence, that the second
order susceptibility $\chi ^{(c,2)}$ is well fitted to 
uncover cooperative effects and to measure the coherence length
in disordered and glassy systems. Importantly, this susceptibility 
has a simple operative definition, which might be fitted to experimental
investigations.
Finally, we mention that
the relevance of the beyond-linear FDR 
is not restricted to the issue of cooperativity, but is related to a number 
of open questions among which the extension of the 
concept of effective temperatures beyond linear order.


\begin{thebibliography}{99}
\bibitem{nota2}
More details and higher order calculations will be presented elsewhere.
\bibitem{nota5}
The FDR, 
in the context of the
Langevin equation, has been derived also in~\cite{biroli}.
However, there are discrepancies between the results in~\cite{biroli} and ours. 
The main ones are i) in the prefactors
involving $\beta$ and ii) in the contention made in~\cite{biroli}, which we do not agree with,
that the correlation functions with $B$ appearing with the shortest
time do vanish in equilibrium (see discussion below Eq.~(\ref{chiscal})).
\end{thebibliography}
\end{document}